# Design and Implementation of a Simple Web Search Engine

Andri Mirzal

*Faculty of Computer Science and Information Systems,
University of Technology Malaysia
andrimirzal@utm.my*

*Abstract*

*We present a simple web search engine for indexing and searching html documents using python programming language. Because python is well known for its simple syntax and strong support for main operating systems, we hope it will be beneficial for learning information retrieval techniques, especially web search engine technology.*

***Keywords:*** *hyperlink structure, python programming language, ranking algorithms, search engine, web crawler*

## 1. Introduction

Many papers written in the web information retrieval (IR) field utilize their own web crawlers to crawl, index, and analyze contents (including hyperlink texts) of the pages and network structure of the web. Sometimes a search function to return relevant pages to the users' queries is also provided. Crawler and search function are considered to be the fundamental components of a search engine [1], and each has its own research challenges and problems.

Web crawler, also known as spider or robot, is responsible for fetching pages, parsing hyperlinks, managing crawl queue, and indexing contents of the pages. In a more sophisticated form, this component also implements politeness policies (e.g., obeying robot.txt instructions and not overloading servers with repetitive pages queries [1, 2, 3]), indexes anchor text [4], and decides in advanced which are spam links or pages with different URLs but similar contents (so that the crawler can avoid downloading these pages to save bandwidth [1, 5, 6]). Some of the real working crawlers like IRLbot [1], Mercator [2, 7], Polybot [3], iRobot [6], UbiCrawler [8], and Googlebot [9] provide good documentations on their designs and implementations. However, the descriptions are still too general to make any reproduction effort possible.

Search function is an interface between a search engine and users. This function receives users' queries and returns relevant pages to the queries. The pages usually are sorted according to some criteria. The most basic criterion is Boolean match between contents of the pages and words in the queries. More advanced mechanisms use hyperlink structure of the web graph (e.g., PageRank [9], HITS [10], and Salsa [11]), anchor text information [12, 13, 14, 15], and user-click behavior [4, 16] to determine ranking of the relevant pages.

There are some notable open source search engine projects like Lemur (lemurproject.org) and Lucene (lucene.apache.org). But due to their complex design and implementation, these projects are still not a good starting point to learn about web search engine technology. Here the importance of open source search engine project that the design and implementation are easy to understand is emphasized because there are many occasions that we want to do something different than those that already have been done, e.g., implementing new ranking algorithms, doing stemming on contents and hyperlinks texts, and using new methods in anchor text analysis. All these require us to alter the existing codes and add some new functionality to the codes.

To provide software that satisfies the above conditions, we propose using search engine program written by Toby Segaran [4]. This program is written in python, a scripting language that has been very popular recently due to its simple syntax and comprehensive libraries. The program itself comprises of three classes: `crawler`, `searcher` and `nn` (neural network)[1]. Here, we only use, improve, and add new functionalities to the `crawler` and `searcher` classes because we are only interested in building search engine to crawl web pages and provide search function to the users, not in training neural network with users' feedback which requires log dataset. Readers who are interested in implementing neural network can consult the book [4] and a paper by Baeza-Yates et. al. [16] in which they use search log information to classify users' queries.

## 2. System Overview

The system is designed and implemented by pertaining the original architecture. In the top level view, the system consists of crawler, searcher, and database components. Figure 1 shows overall design of the system, figure 2 shows more details of the system design, and figure 3 and 4 show structures of the `crawler` and `searcher` classes respectively. The following subsections describe the modifications to the original system.

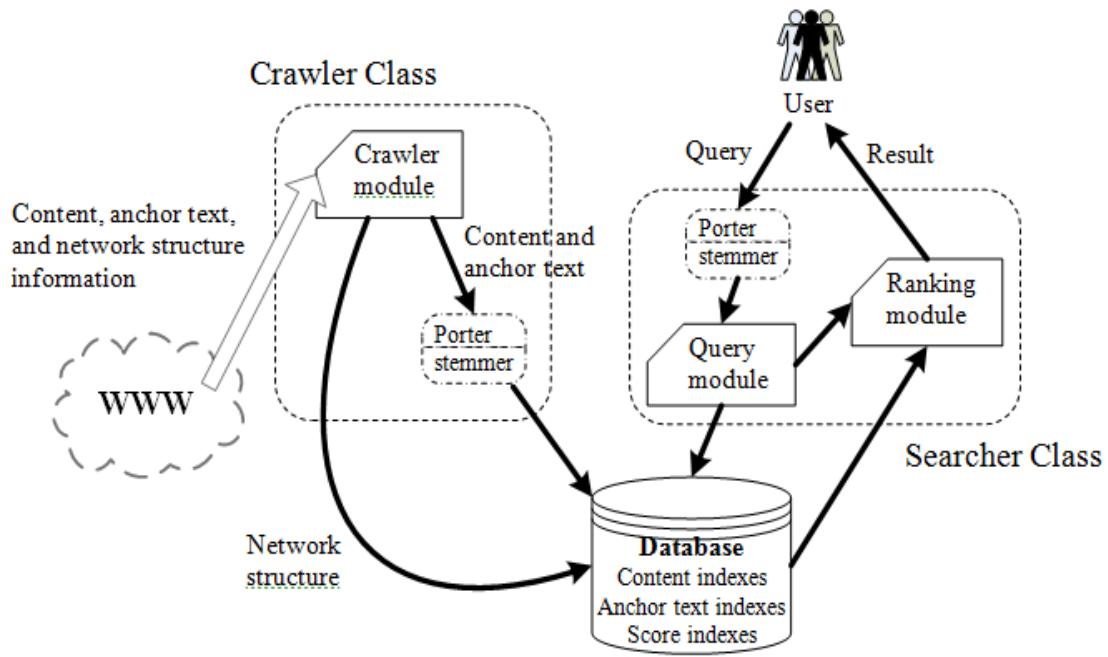

**Figure 1. System Design in Top Level View**

---

[1] http://kiwitobes.com/PCI_Code.zip

## 2.1. Improve Reliability of the Crawling Process

In the original code, the crawler will attempt to open a URL randomly in the URLs list (stored as a python set) provided by the user. If it fails, the crawler will start with the next URL. The problem with this strategy is even though the URL can be opened, it doesn't mean that the content can be read. And further if the content can be read, sometimes the indexer functions in the `crawler` class can fail to input the indexes into corresponding tables in the database. In many cases, these errors can cause the program to stop working, which is very inconvenient because in web data mining the ability to continue crawling process by ignoring the errors is an indispensable quality.

We overcome this shortcoming by simply introducing additional `try-except` statements for opening & reading URLs and indexing purposes. With this simple approach we were able to crawl web pages for three days without stopping and downloaded 74,243 pages from scholarpedia.org using notebook with 1.86 GHz Intel processor, 2 GB RAM, and 100 Mbps (effective about 53 Mbps) Internet connection before the process terminated. Compared to the original system, where the crawling will end immediately if there is an error, this is a significant improvement (the final version of our system, available at pythinsearch.googlepages.com, could crawl the web pages for about two months continuously when used to collect datasets for testing our algorithms in ref. [17]).

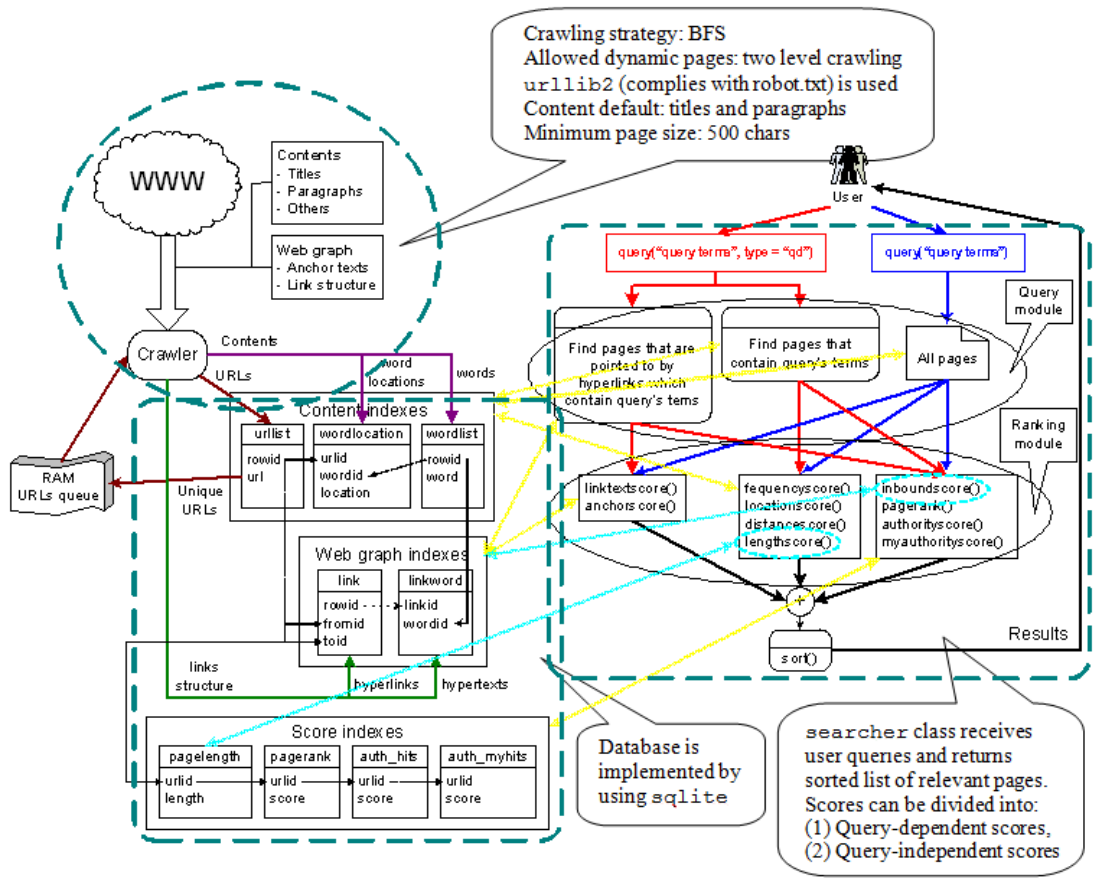

**Figure 2. More Details on System Design**

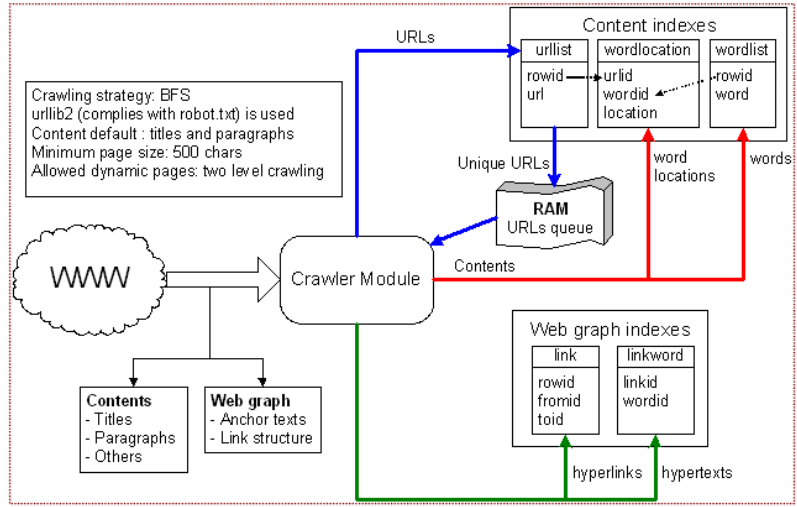

**Figure 3. The `crawler` Class Structure**

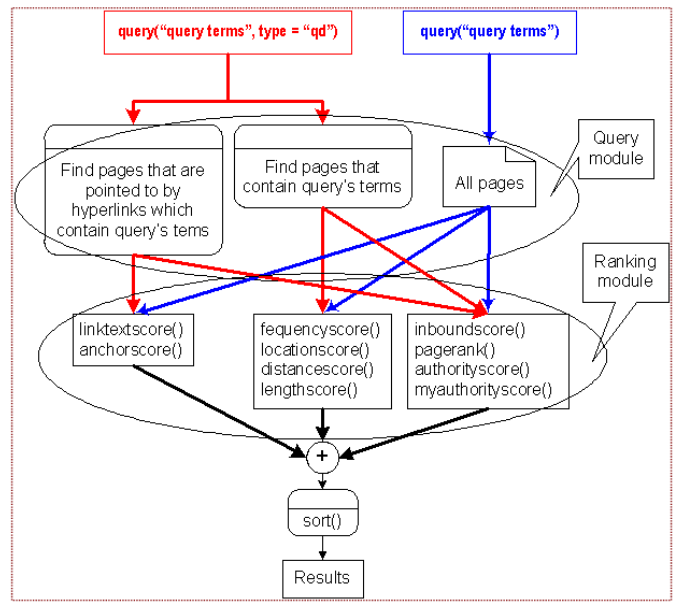

**Figure 4. The `searcher` Class Structure**

We also add stop words according to English standard stop words[2] to reduce the size of database. But 'computer' term is eliminated from the list because we believe this term is meaningful in the search purpose.

## 2.2. Add an Optional Stemming Function

We add stemming function using porter stemmer algorithm [18] to both the `crawler` and `searcher` classes. Stemming is the process to reduce inflected and derived words into their stems. This process frequently improves the performance of an IR system, because words with the same stem usually have similar meaning. The stems itself need not to be the same with their morphological roots, but a good stemming algorithm should return the same stem for words that have the same morphological root. For example 'navigational', 'navigation', and 'navigate' have morphological root 'navigate'. If porter stemmer is used, 'navig' will be returned instead of 'navigate' for all these words.

---

[2] http://www.dcs.gla.ac.uk/idom/ir_resources/linguistic_utils/stop_words

As shown in figure 1, the stemming function is located before content and anchor text enter the database in the `crawler` class and before queries enter query module in the `searcher` class. This procedure will ensure that all words in `wordlist` table (see figure 5 (a)) being saved in their corresponding stems, and all terms in queries will enter the query module also in their respective stems. One important thing to remember is, if crawling process conducted with stemming function activated in the `crawler` class, stemming function in the `searcher` class must also be activated.

### 2.3. Add New Scores Functions

We incorporate some standard scores functions that haven't available in the original code. The first function is `bm25score()` that uses Okapi bm25 algorithm [19], an algorithm that calculates the frequencies of appearance of query terms in the pages with assumption that each term is independent. This algorithm is the standard algorithm in content-based document retrieval research and seems to be the default retrieval function in IR research community.

The second function is `calculatelength()` that creates `pagelength` table with `urlid` as the primary key and `length` of the pages as the return value. This function scores pages based on their lengths where longer pages have higher scores than the shorter ones. The function that accesses `pagelength` table and returns length scores is `lengthscore()`. Note that this function not only returns pages that contain terms in query, but also pages that being linked by anchor texts which contain terms in query. This approach is used because anchor texts are proven to be good indicators about pages' contents and behave as "consensus titles" [20].

The third is `calculatehits()` that calculates authority and hub scores of each pages based on query-independent HITS [21 pp. 124—126], an algorithm that computes global authority and hub vectors which consequently slightly reduce the influence of link spamming. This function creates two tables, `auth_hits` and `hub_hits` for each `urlid`. Function `authorityscore()` accesses `auth_hits` table and returns authority scores. Curious readers can also write a function similar to `authorityscore()` that returns hub scores by only change `auth_hits` with `hub_hits` statement (in real application, hub scores are rarely used to rank pages).

The fourth function, `calculatemyhits()` is based on our proposed ranking algorithm. This function calculates authority and hub scores of each page using algorithm described in ref. [17], and stores the results in `auth_myhits` and `hub_myhits` table. Function `myauthorityscore()` then is used to read scores from `auth_myhits` table.

The last added score function is `anchorscore()`, a function with four optional schemes that returns relevant pages according to the frequencies of appearance of query's terms in the anchor texts linking to the pages. This is a new method and has been proven to be very effective in finding the relevant pages, especially if the anchor model (scheme 4, the default scheme) is used. The detail discussion and performance evaluation can be found in ref. [15].

### 2.4. Organize the Scores Functions into Query-dependent and Query-independent

Some scores depend on queries and some are independent so can be calculated in advanced and stored in database. To simplify, it can be considered that query-dependent scores are content-based or anchor text-based scores, and query-independent scores are link structure-based scores. However, `lengthscore()` is an exception since it is a content-based scores but query-independent.

The query-dependent functions can be changed into query-independent by passing string argument other than 'qd' into `query()` function in the `crawler` class. The `query()` itself receives two arguments; the first is the query (string) and the second is the type of return scores, whether it is query-dependent or query-independent. The default value is 'qd' (query-dependent). See readme.txt on the source code[3] for more information.

### 2.5. Add Additional Indexes to `linkwords` Table

Because four additional schemes to calculate scores based on anchor text [15] are implemented, it is necessary to create new indexes on `linkwords` table (each for `linkid` and `wordid` columns) to boost performances of scores functions that call entries in `linkid` and `wordid`. By creating these indexes, the execution time for `linkwords` related scores functions, `linktextscore()` and `anchorscore()`, becomes 3-4 times faster.

## 3. Database Design

There are 12 tables in the database which divided into three categories, content, web graph, and score indexes tables. The shaded tables are the added tables. The content indexes tables are mainly used in the content-based scores functions, web graph indexes tables are mainly used in the link structure-based and anchor text-based scores functions, and score indexes tables are the tables that store query-independent scores. Query-dependent scores, as noted earlier, cannot be calculated in advanced, so they cannot be stored in tables.

The `urllist` table stores the crawled web addresses in consecutive numbers based on the time they were crawled. These numbers, `rowids`, become identity numbers for the corresponding web addresses. All other tables that have to utilize web addresses use the `rowids` instead of directly using URL names. The arrows from `urllist` to entries in others tables indicate that those entries are pointers to `urls` by using `rowids` of `urllist` table.

The same strategy is also applied in `wordlist` table, where its `rowids` become identity numbers for corresponding words. And then `wordlocation` table utilizes both `urllist` and `wordlist` to make a list of locations of all words in every page. The `location` itself is simply the order of corresponding word appearance on the pages.

The `link` table stores network structure of the crawled web pages. It is the table that the link structure ranking algorithms use to calculate pages' scores. The `linkword` table stores anchor texts of the corresponding links which are used in the anchor text analysis and scoring.

All tables in score indexes simply store the scores of each page in the collection. When users input queries, search function will call appropriate scores functions that utilize these tables. The third entries in `auth_myhits`, `hub_myhits`, and `mypagerank` are constants defined in our proposed algorithms calculated based on number of inlinks and outlinks of each page. See [17] and [22] for more detailed discussion on these algorithms.

---

[3] pythinsearch.googlepages.com

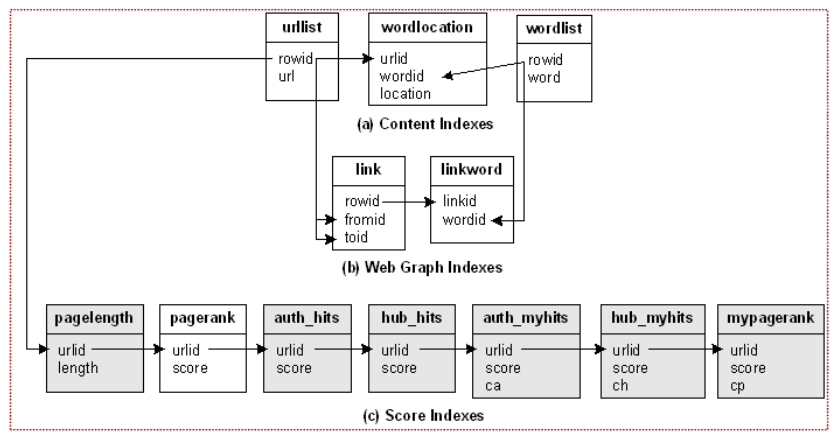

**Figure 5. Database Structure**

## 4. Conclusions

We have presented web search engine software that is suitable for researches and learning purposes because of its simplicity, portability, and modifiability. The strength of our program is in the search function component since we provided many scores functions to sort relevant pages to user queries; especially, the inclusion of anchor text analysis makes our program can also find relevant pages that do not contain terms in the queries.

In the crawler component, only small modification was made. However, this small modification can improve the crawling reliability significantly. Readers who have read the documentation would notice that the crawling method is breadth first search without politeness policies (e.g., obeying robot.txt and controlling access to the servers), spam pages detection, priority URLs queue, and memory management to divide the load of crawling process between disk and RAM. Without good memory management, all of URLs seen tasks are conducted by searching in `urllist` table in the disk which is very time consuming. We will address the crawler design problem in the future researches.

Because there is no single best way to choose the combination between scores functions and their weights, we encourage users to experiment with their own databases. But perhaps some guidelines here can be considered. For example if database contains a set of pages that linked by very descriptive anchor texts, like documents from online encyclopedia, then `linktextscore()` and `anchorscore()` can be given relatively strong weights. And because authors can easily create pages with many occurrences of keywords, `frequencyscore()` and `bm25score()` will give misleading results in this situation, so their weights must be set smaller compare to more reliable content-based metric, `locationscore()`.

## Author

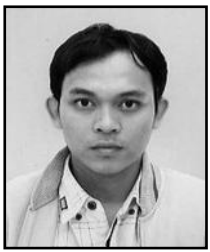

**Andri Mirzal** received PhD and MSc in Information Science and Technology from Hokkaido University, and Bachelor of Engineering in Electrical Engineering from Bandung Institute of Technology. His research interests include machine learning, computational intelligence, and web search engine. Currently, he is a Senior Lecturer at Faculty of Computer Science and Information Systems, University of Technology Malaysia.